# INFORMATION RETRIEVAL IN INTELLIGENT SYSTEMS:
# CURRENT SCENARIO & ISSUES

Sudhir Ahuja[1], Mr. Rinkaj Goyal[2]

[1]University School of Information Technology,
Guru Gobind Singh Indraprastha University,
New Delhi, India
Email: sudhir_ahuja2002@yahoo.com

[2]University School of Information Technology,
Guru Gobind Singh Indraprastha University,
New Delhi, India
Email: rinkajgoyal@gmail.com

**Abstract:** Web space is the huge repository of data. Everyday lots of new information get added to this web space. The more the information, more is demand for tools to access that information. Answering users' queries about the online information intelligently is one of the great challenges in information retrieval in intelligent systems.

In this paper, we will start with the brief introduction on information retrieval and intelligent systems and explain how swoogle, the semantic search engine, uses its algorithms and techniques to search for the desired contents in the web. We then continue with the clustering technique that is used to group the similar things together and discuss the machine learning technique called Self-organizing maps [6] or SOM, which is a data visualization technique that reduces the dimensions of data through the use of self-organizing neural networks. We then discuss how SOM is used to visualize the contents of the data, by following some lines of algorithm, in the form of maps. So, we could say that websites or machines can be used to retrieve the information that what exactly users want from them.

**Keywords:** SOM, SWOOGLE, self-organizing maps, Intelligent Information Retrieval, clustering.

## 1. Introduction

Intelligent Information Retrieval (IR) [8] has been defined by different people in various manners, but the consistent theme has been one of the machine (or program) doing something *for* the user, or the machine (or program) *taking over some* functions that previously had to be performed by human beings (either user or intermediary). Information Retrieval Tool [1] is an important way for people to obtain knowledge and information. However, the technique of traditional information retrieval has been criticized as deeply flawed; the key reason is that the search technique is mainly based on the keyword match. In other words, users enter the inputs in the form of keywords which they want to search, and then retrieval system, like search engines, return the matching documents to users. Because of polysemy and synonyms, it is very hard to understand user's exact requirements by keywords. Often the keywords entered do not get the results they want, the somewhat relevant or irrelevant documents are also retrieved.

Let's say that a user starts using a web application to order food, and he order fish every Sunday. He has much better understanding if, on Sundays, the application asked him "What would you like to order today?" rather than "Would you like to have fish today?" In the second case, the application somewhat realised that he likes fish on Sundays. Thus, the data created by user's interaction with the site doesn't affect how the application chooses the content of a page or how it's presented. Asking a question that's based on the user's prior selections introduces a new kind of interactivity between the website and its users. So, we could say that websites with that property have a *learning capacity* [2].

*The Information Retrieval [18] system consists of three major components* [8]:

1. the *user(s)* in the system;

2. the *knowledge resource* to which the user has access and with which he/she interacts; and,

3. Some person(s) and/or device(s) which supports and mediates the user's interactions with the knowledge resource (the *intermediary*).

*Limitations in the past system*

1. It is difficult to understand needs of the user so as to search for the specified sets of keywords.

2. Traditional websites or machines do not have the ability to learn and were not capable enough to assume the search criteria; and hence were not able to find the results based on the inputs entered by the users in the past.

*Advantage in the current system*

We aim to build machines that can record all the past inputs and behaviors entered by the users. The machine should be able to group the similar things together and represent the information in the form of graphs. The tool we will use can visualize the information of high dimensional data by reducing its dimensions.

One advantage [9] of the use of SOM with respect to other clustering algorithms is the spatial organization of the feature map that is achieved after the learning process.

Basically, more similar clusters are closer than more different ones.

## 2. Basic elements of intelligent applications [2]

> *2.1 Aggregated content* — Large collection of data similar to or dependent on particular application. The content that is aggregated is dynamic rather than static, and its origins as well as its storage locations could be geographically dispersed. Each and every piece of information is associated or linked with many other different pieces of information.
>
> *2.2 Reference structures*: These are the structures that provide one or more structural and semantic interpretations of the content. For example, this is related to what people call *folksonomy [2]* — the use of tags for annotating content in a dynamic way and continuously updating the representation of the collective knowledge to the users.

2.3 *Algorithms*: The algorithms are applied on the aggregated content, and sometimes require the presence of reference structures.

The above elements, summarized in figure below, are required for making an application to an intelligent application.

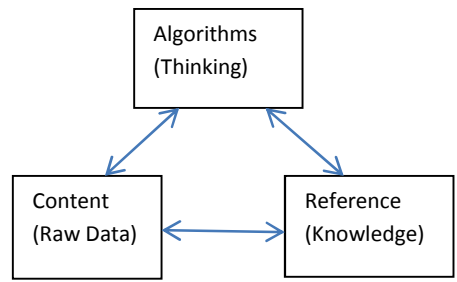

Fig 1: Elements of intelligent applications

*How can I build intelligence in my own application? [2]*

- *Examine your functionality and your data*

Use cases should be identified, so as to get the behavior of the intelligent application, at the start of the process. These cases can be identified by asking a set of questions. Some of such questions could be like:

1. Does our application have the good content that is collected from various sources?

2. Does our application is a web application or a machine dependent?

3. Does the application deal with free text?

4. Does the application have wizard-based workflows?

5. Does our application provide search functionality?

6. Does the application require any kind of reporting?

7. Does our application is capable to make automated decisions based on defined rules?

8. Is identity verification important for our application?

9. Does our application is capable enough to recommend the search results?

The above list is just the set of some example questions that may be identified in use cases, as these questions will help us to provide indications of the possibilities. If the answer to any of such questions is yes, our application can benefit greatly from the techniques like clustering, etc.

- *Get more data from the web*

Our own data will not be sufficient, in some cases, for building intelligence to our application. This may require access to external information from web.

*Eight fallacies of intelligent applications [2]*

1. Data is reliable
2. Data size does not matter
3. Solution's scalability is not an issue
4. Apply the same good library everywhere
5. Complicated models are better
6. The computation time is known
7. There are models without bias
8. Inference happens instantaneously

## 3.  Some Techniques for retrieving intelligent information:

### 3.1 SWOOGLE [10]: An engine for semantic web

Swoogle [15], the Semantic web search engine, is a research project carried out by the ebiquity research group in the Computer Science and Electrical Engineering Department at the University of Maryland. It's an engine aimed towards finding documents on the semantic web.
Swoogle [3] is a crawler-based indexing and retrieval system for the Semantic Web. It extracts metadata for each discovered document, and computes relations between documents.

### 3.2 Clustering

The term *clustering* [2] means that the process of grouping similar things together. Clustering [12] is suitable in many situations, where the grouping for similar items or things is required, but it is not always possible to achieve a desired aim by simply issuing SQL queries. In many cases, the search elements that we need to use for locating the desired results are not the unique identifiers, so we need to use techniques that work well with arbitrary data.

Clustering algorithms [9] are well suited to deal with unlabeled patterns and this is particularly important in applications where the human validation of the pattern membership can be very expensive.

Supervised vs. Unsupervised Learning

1. An important area of an Artificial neural network mode is whether it needs assistance in learning or not. Based on the way of their learning, all artificial neural networks can be categorized into two learning types - supervised and unsupervised.

2. In supervised learning, an ideal output result for every single input vector is essential when the network is trained. It is thus possible to make the neural network learn the behavior of the process under study.

3. In unsupervised learning, the training of the network is totally data-driven and there are no target results for the input data vectors provided. An Artificial neural network of the unsupervised learning category, such as the self-organizing maps (SOM) [14], can be used for clustering the data and find similarities and features inherent to the problem.

**Self-Organizing Map**

Self-organizing map [6] is a technique invented by Professor Teuvo Kohonen and it generally refers to Kohonen's Self Organizing Map [11], or SOM for short. It is a data visualization technique which is used to reduce the dimensions of data through the use of self-organizing neural networks.

SOMs are used to accomplish two tasks, they reduces the high dimensions of the data and displays the grouping of all the available similarities among that data by producing a map of mostly dimensions 1 or 2.

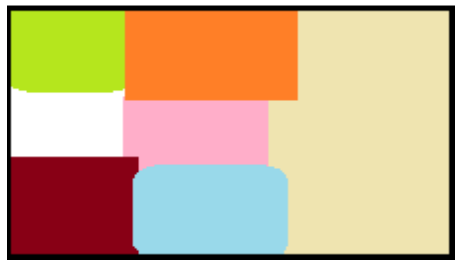

Fig 2: Two dimensional image as a result of SOM

The figure above gives an idea of a 2 dimensional image similar to one that can be created using any one of the available SOM tools. The figure shows the grouping of the colors like all the greens are shown on the upper left corner and in the same way all the blues are shown in the other side of the image.

Components [6]

- *Sample Data*

Sample data is the first part of a SOM. The figure below states some examples of 3 dimensional data which are usually used when experimenting with SOMs. In the below figure, the colors are represented in three dimensions (blue, red and green). The basic idea of the self-organizing maps (SOM) is to display and project the n-dimensional data (in this example it would be colors and would be 3 dimensional) into something that can be better understood visually.

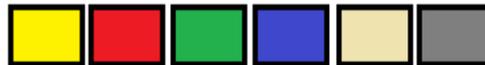

Fig. 3: Sample Data

- *Weights*

Weight vectors are the second component of SOMs. Each weight vector has its corresponding two components to which we have here attempted to show in the figure below. Data is the first part of weight vector, which is of the same dimensions as the sample vectors; and natural location is the second part of a weight vector. In this case, color is the data and location is (x,y) position of the pixel on the screen which is a good thing about data as colors are easy to identify.

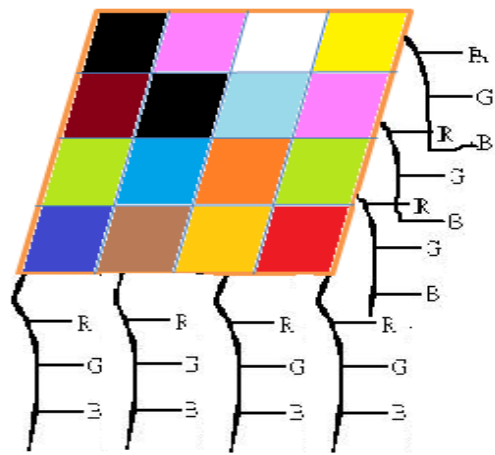

Fig 4: 2D Array Weight of Vector

In the previous example, 2D array of weight vectors were used and would look like figure above. The picture above is a skewed view of a grid, where we have the n-dimensional array for each weight and each weight has its own unique location in the grid. Weight vectors do not need to be arranged in 2 dimensions, a lot of work has been done using SOMs of one dimension, but the data part of the weight needs to be of the same dimensions as the sample vectors. Weights are sometimes mentioned as neurons [2] as SOMs are actually neural networks.

## SOM Algorithm:

Due to the very nature of self-organizing maps, the way they go about organizing themselves is by competing for representation of the samples. Neurons are also permitted to change themselves by learning so that to become more like samples in the hope of winning the next competition. It is this learning and selection process that makes the weights organizes themselves into a map of representing similarities.

SOM algorithm [7] is used, with the use of two components (the sample and weight vectors), that will represent the similarities of the sample vectors.

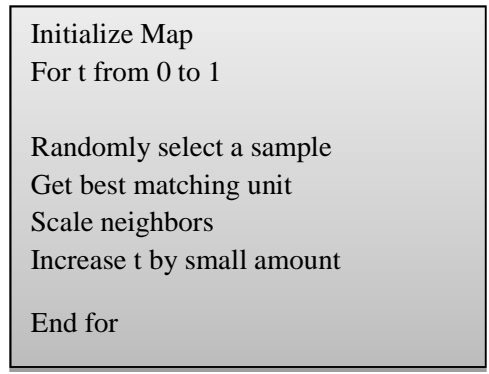

Fig 4: SOM Algorithm

The algorithm starts with the initialization of the weight vectors. After doing this, select the sample vectors randomly from the available weight vectors and search the weight vectors in the map to find which one best match that sample. Next step is to find the neighbors of the selected weight vectors that are close to it. By doing this, the neighbors are rewarded with the like of randomly selected vector. After doing this, algorithm increases the value of it by some small amount and the procedure for randomly selecting the weight vectors and finding matches with the neighbors continues for several numbers of times.

Advantages of using SOM over other clustering techniques:

1. *Easy to understand* [7]: SOMs are easy to understand and quite simple to analyze the information contained in it. The color combinations help to easily visualize the generated map.

2. *Works very well* [7]: As shown previously, SOMs classify data well and are able to easily evaluate for their own quality. So one can actually evaluate how good a map is and how strong the relationships between objects are.

Issues while using SOM [7]

1. *Getting the correct data* [7]: One problem with SOMs is getting the right and correct data. This is because we need a value for each dimension of each member of samples in order to generate a map. Sometimes, this may not be easily possible and it becomes very difficult to gather all of this data. So this is a major disadvantage of using them.

2. *Different similarities among different SOMs* [7]: Due to availability of many SOMs, every other SOM evaluates different similarities among the sample vectors. SOMs organize samples so as to gather all the similar data surrounded to the sampled data. There may be cases when similar samples are not near to each other. So lots of map needs to be constructed in order to get one final good map.

3. *Computationally expensive* [7]: Sometimes due to increase in the dimensions of the data, the need for dimension reduction visualization techniques becomes important. This leads to the more time for computation.

## 4. Conclusion & Future work

We have discussed the techniques for intelligent information retrieval and machine learning using clustering and self-organizing map. In future we can implement this machine learning ability using one algorithm that introduces the vital concepts of distance and similarity. This will present techniques for generating similar data together.

## 5. References

1. Building Intelligent Information Retrieval System Based on Ontology, "Pan Ying Wang Tianjiang Jiang Xueling", The Eighth International Conference on Electronic Measurement and Instruments, ICEMI' 2007.
2. Algorithms of the Intelligent Web, "Haralambos Marmanis, Dmitry Babenko" Manning, Greenwich, May 2009.
3. Swoogle: A Semantic Web Search and Metadata Engine, "Li Ding, Tim Finin", Department of Computer Science and Electronic Engineering, University of Maryland Baltimore County, Baltimore MD 21250, USA.
4. Using Software Agents for Information Retrieval in a semantic web environment, Teresa Herrmann, July 18, 2006.
5. Intelligent web traffic mining and analysis, "Xiaozhe Wang, Ajith Abraham, Kate A. Smith", 7 January 2004.
6. Self Organising Maps, "Tom Germano", March 23, 1999.
7. Types of Machine Learning Algorithms, "Taiwo Oladipupo Ayodele", *University of Portsmouth, United Kingdom.*
8. Intelligent Information Retrieval: Whose Intelligence? "Nicholas J. Belkin", "School of Communication, Information and Library Studies", Rutgers University, 4 Huntington Street, New Brunswick, NJ 08901-1071 USA.
9. SOM clustering for text retrieval and classification with examples on Indian scripts, "Simone Marinai", Dipartimento di Sistemi e Informatica - Universit`a di Firenze.
10. Document Ranking on Semantic Web, "Asmita Rahman and Shasha Liu", Advance Topics in Information Systems, Term Project Proposal, Feb 2010.
11. Self-Organizing Maps, "Kohonen T", 3rd ed., Information Sciences. Berlin, Heidelberg, Springer, 2001.
12. Application of Self-Organizing Maps to the Maritime Environment, "Victor J.A.S. Lobo", Portuguese Naval Academy, Almada, Portugal.
13. Introduction to Machine Learning, "Taiwo Oladipupo Ayodele", University of Portsmouth, United Kingdom.
14. Self-Organizing Maps for Clustering in Document Image Analysis, "Simone Marinai, Emanuele Marino, and Giovanni Soda", University of Florence Dipartimento di Sistemi e Informatica (DSI), Firenze, Italy.
15. *Swoogle: Searching for knowledge on the Semantic Web, "Tim Finin, Li Ding, Rong Pan, Anupam Joshi, Pranam Kolari, Akshay Java and Yun Peng", University of Maryland Baltimore County, Baltimore, MD.*
16. Information Retrieval Tools and Techniques, "A. Accomazzi", Smithsonian Astrophysical Observatory, Cambridge, MA, USA.
17. Knowledge Resource Tools for accessing large text files, "Donald E. Walker", Conference on Theoretical and Methodological Issues in Machine Translation of Natural Languages, Colgate University, Hamilton, New York.
18. Info Crystal: A Visual Tool for Information Retrieval, "Anselm Spoerri", Department of Civil and environmental Engineering, Massachusetts Institute of Technology, February 1995.